\documentclass[twocolumn,showpacs,preprintnumbers,amsmath,amssymb]{revtex4}
\usepackage{graphicx}
\usepackage{dcolumn}
\usepackage{bm}

\def\1{\mathchoice{\rm 1\mskip-4.2mu l}{\rm 1\mskip-4.2mu l}{\rm
        1\mskip-4.6mu l}{\rm 1\mskip-5.2mu l}}

\begin{document}
\title{Observation of nonadditive mixed state phases with\\polarized neutrons}
\author{J\"{u}rgen Klepp$^{1}$,
Stephan Sponar$^{1}$,
Stefan Filipp$^{1}$,
Matthias Lettner$^{1}$,
Gerald Badurek$^{1}$
and Yuji Hasegawa$^{1,2}$}
\affiliation{$^1$Vienna University of Technology, Atominstitut, Stadionallee 2, 1020 Vienna, Austria}
\affiliation{$^2$PRESTO, Japan Science and Technology Agency, 4-1-8 Honcho Kawaguchi, Saitama, Japan}
\date{\today}
\begin{abstract}
In a neutron polarimetry experiment the mixed state
relative phases between spin eigenstates are determined from the maxima and minima of measured intensity oscillations. 
We consider evolutions leading to
purely geometric, purely dynamical and combined phases. 
It is experimentally demonstrated that the sum of the 
individually determined geometric and dynamical phases 
is not equal to the associated total phase which is 
obtained from 
a single 
measurement, unless the system is in a pure state.
\end{abstract}
 \maketitle 
Evolving quantum
systems acquire two kinds of phase factors: (\emph{i}) the
dynamical phase which depends on the dynamical properties of the
system - like energy or time - during a particular evolution, and
(\emph{ii}) the geometric phase which only depends on the path the 
system takes in state space on its way from the
initial to the final state. Since its discovery by Pancharatnam
\cite{Pancharatnam1956} and Berry \cite{Berry1984} the concept was
widely expanded and has undergone several generalizations.
Nonadiabatic
\cite{AharanovAnandan1987} and noncyclic
\cite{SamuelBhandari1988} evolutions as well as the off-diagonal
case, where initial and final state are mutually orthogonal
\cite{ManiniPistolesi2000}, have been considered. Ever since, a
great variety of experimental demonstrations has been accomplished \cite{TomitaChao1986,SuterEtAl1988},
also in neutron optics
\cite{BitterDubbers1987,AllmanEtAl1997,HasegawaEtAl2001,FilippEtAl2005}.
Due to its potential robustness against noise 
\cite{LeekEtAl2008} the geometric phase is an excellent
candidate to be utilized for logic gate operations in 
quantum information science \cite{NielsenChuang}. 
Thus, a rigorous investigation of all its properties 
is of great importance. 
\newline\indent In addition to an approach by Uhlmann
\cite{Uhlmann1991} a new concept of phase for mixed 
input states based on interferometry was developed by 
Sj\"{o}qvist \emph{et al.} \cite{SjoeqvistEtAl2000}. 
Here, each eigenvector of the initial
density matrix independently acquires a geometric phase. 
The total mixed state phase is a weighted average of the 
individual phase factors. 
This concept is of great significance for such
experimental situations or technical applications 
where pure state theories may imply strong 
idealizations. Theoretical predictions have 
been tested by Du
\emph{et al.} \cite{DuEtAl2003} and Ericsson \emph{et al.}
\cite{EricssonEtAl2004} using NMR and single-photon
interferometry, respectively. 
In this letter, we report on measurements of 
nonadiabatic and noncyclic geometric, dynamical
and combined phases. These depend on noise 
strength in state preparation, defining the degree 
of polarization, the purity, of the neutron input 
state.
In particular, our experiment demonstrates that the 
geometric and dynamical 
mixed state phases $\Phi_{\mbox{\scriptsize{g}}}$ and $\Phi_{\mbox{\scriptsize{d}}}$, 
resulting from separate measurements, are not additive 
\cite{SinghEtAl2003}, because the phase resulting from a 
single, cumulative, measurement differs from  $\Phi_{\mbox{\scriptsize{g}}}+\Phi_{\mbox{\scriptsize{d}}}$.
This nonadditivity might be of
practical importance for applications, e.g. the design of 
phase gates for quantum
computation. \linebreak\indent
A neutron beam propagating in $y$-direction interacting with static magnetic fields $\vec {\mbox{B}}(y)$
is described by the Hamiltonian H$=-\hbar^2/2m\vec\nabla^2-\mu\vec\sigma\vec B(y)$;
$m$ and $\mu$ are the mass and the magnetic moment of the
neutron, respectively. Zeeman splitting within $\vec
{\mbox{B}}(y)$ leads to solutions of the Schr\"{o}dinger equation
$\cos(\theta/2)|k_+\rangle|+\rangle+
e^{i\alpha}\sin(\theta/2)|k_-\rangle|-\rangle$,
where $|k_{\pm}\rangle$ are the momentum and $|\pm\rangle$ the
spin eigenstates within the field $\vec {\mbox{B}}(y)$. $\theta$
and $\alpha$ denote the polar and azimuthal angles determining
the direction of the spin with respect to 
$\vec {\mbox{B}}(y)$.
$k_{\pm}\simeq k_0\mp\Delta k$, where $k_0$ is the momentum 
of the free particle and 
$\Delta k=m\mu|\vec{\mbox{B}}(y)|/\hbar^2k$ is
the field-induced momentum shift. $\Delta k$ can
be detected from spinor precession. 
Omitting the coupling of momentum and spin, we focus on the
evolution of superposed spin eigenstates 
resulting in Larmor
precession of the polarization vector
$\vec r=\langle\varphi|\vec\sigma|\varphi\rangle$, where $\vec\sigma=(\sigma_x,\sigma_y,\sigma_z)$ is the Pauli vector operator.
\begin{figure}
    \scalebox{0.31}{\includegraphics {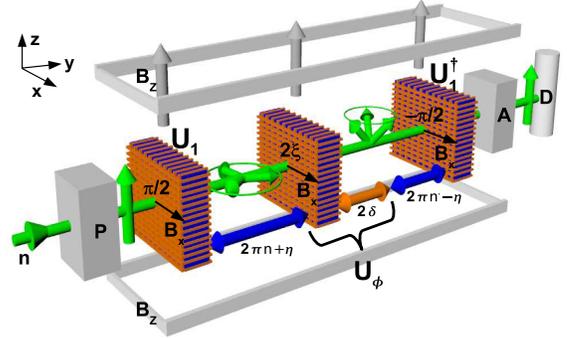}}
    \caption{Sketch of neutron polarimetry setup for phase measurement with overall guide field B$_z$, polarizer P,
    three DC-coils to implement unitary operations U$_1$, U$^{\dagger}_1$, U$_\phi$, analyzer A and detector D.
    Greek letters denote spin rotation angles. Shifting the second coil induces an additional dynamical phase $\eta$ resulting in intensity 
    oscillations. The desired phase $\phi$ is determined from their minima and maxima.}
    \label{fig1}
\end{figure}
The unitary, unimodular operator
\begin{eqnarray*}
    \mbox{U}(\xi',\delta',\zeta')&=&e^{i\delta'}\cos\xi'|+\rangle\langle +|-e^{-i\zeta'}\sin\xi'|+\rangle\langle -|\\
    &+&e^{i\zeta'}\sin\xi'|-\rangle\langle +|+e^{-i\delta'}\cos\xi'|-\rangle\langle -|
\end{eqnarray*}
describes the
evolution of the system within static magnetic fields. The set of SU(2) parameters
$(\xi',\delta',\zeta')$ is related to the
so-called Cayley-Klein parameters $a,b$ via
$a=e^{i\delta'}\cos\xi'$ and $b=-e^{-i\zeta'}\sin\xi'$ (see e.g.
\cite{Sakurai}).
\begin{figure}
    \scalebox{0.15}{\includegraphics{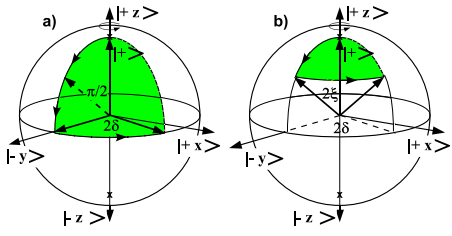}}
    \caption{Evolution of the $|+\rangle$ eigenstate 
				induced by U$_\phi$, associated to:
            a) Purely (noncyclic) geometric phase ($2\xi=\pi/2$).
            b) Combinations of dynamical and geometric phase on the Bloch sphere ($0<2\xi<\pi/2$).}
    \label{fig2}
\end{figure}

Consider the experimental setup shown in Fig.\,\ref{fig1}. 
A monochromatic neutron beam passes the polarizer P 
preparing it in the 'up' state $|+\rangle$ with
respect to a magnetic guide field in $z$-direction (B$_z$). 
Next, the beam approaches a DC coil with its field 
B$_x$ pointing to the $x$-direction.
B$_x$ is chosen such that it carries out the transformation
U$_1$\,$\equiv$\,U$(\pi/4,0,-\pi/2)$, corresponding to a $+\pi/2$
rotation around the $+x$-axis. After U$_1$ the resulting state of
the system is a coherent superposition of the two orthogonal spin
eigenstates: $|\psi_0\rangle=1/\sqrt{2}(|+\rangle-i|-\rangle)$. 
A subsequent coil, represented by
U$(\xi,0,-\pi/2)$, is set to cause a spin rotation around the
$+x$-axis by an angle $2\xi$. This second coil and the
following propagation distance within B$_z$ --
corresponding to a rotation angle $2\delta$ around 
the $+z$-axis --
define an evolution 
U$_\phi$\,$\equiv$\,U$(\xi,\delta,\zeta)$.
Undergoing the transformation
U$_\phi$ the two spin eigenstates $|\pm\rangle$ acquire
opposite total phase
$\pm\phi
=\pm\arg\langle\pm|\mbox{U}_\phi|\pm\rangle=\pm\delta$.
A third coil (U$^{\dagger}_1$) reverses the action of the first
one and would therefore transform a state $|\psi_0\rangle$ back to
$|+\rangle$. The state of the
system entering the third coil equals $|\psi_0\rangle$ only 
if U$_\phi=\1$. $\phi$ can be extracted by applying an 
extra dynamical phase
shift $\pm \frac{1}{2}\eta$ to $|\pm\rangle$.
It is implemented by adjusting both inter-coil distances 
from first to second and second to third coil to spin 
rotation angle equivalents of $2\pi n+\eta$ and 
$2\pi n'+2\delta-\eta$ ($n,n'$ are integer).
By scanning the position of the second coil these 
rotation angles - referred to as 
U$_{\eta}$\,$\equiv$\,$\mbox{U}(0,\eta/2)$
and U$^{\dagger}_{\eta}$ - are varied 
to yield intensity 
oscillations from which $\phi$ is calculated. 
Note that, because of $\xi'=0$, the parameter $\zeta'$ is 
undetermined and, therefore, omitted in U$_{\eta}$. 
After projection on the state $|+\rangle$ by the 
analyzer A, the phase $\phi\!=\!\delta$
and its visibility $\nu=|\cos\xi|$
can be computed as functions of the maxima and minima of 
the intensity, 
I$_{\mbox{\scriptsize max}}$ and
I$_{\mbox{\scriptsize min}}$, measured by the detector D \cite{WaghRakhecha1995}.
More general, a neutron beam with incident 
purity $r_0'\!=\!|\vec r_0'|$ along the
$+z$-axis ($\vec r_0'\!=\!(0,0,r_0')$) is 
described by the density operator
$\rho_{\mbox{\scriptsize{in}}}(r_0')=1/2(\1+r_0'\sigma_z)$.
Calculating $\rho_{\mbox{\scriptsize{out}}}=\mbox{U}^{\dagger}_1\mbox{U}^{\dagger}_{\eta}
\mbox{U}_\phi\mbox{U}_{\eta}\mbox{U}_1\rho_{\mbox{\scriptsize{in}}}\mbox{U}^{\dagger}_1\mbox{U}^{\dagger}_{\eta}
\mbox{U}^{\dagger}_\phi\mbox{U}_{\eta}\mbox{U}_1 $,
we find the intensity
\begin{eqnarray}\label{mixedstateintensity}
\mbox{I}^{\rho}&\propto& 
\frac{1-r_0'}{2}+r_0'\left(\cos^2\xi\cos^2\delta+\sin^2\xi\cos^2(\zeta-\eta)\right)
\end{eqnarray}
after the analyzer A. Considering the maxima and minima
$\mbox{I}^{\rho}_{\mbox{\scriptsize max}}$,
$\mbox{I}^{\rho}_{\mbox{\scriptsize min}}$ of $\eta$-induced
oscillations of $\mbox{I}^{\rho}$ one obtains the mixed state
phase \cite{LarssonSjoeqvist2003}
\begin{widetext}
\begin{eqnarray}\label{eq:MixedStatePhInTermsOfImaxImin}
    \Phi(r_0')&=&
	\arccos\sqrt{\frac{[\mbox{I}^{\rho}_{\mbox{\scriptsize min}}/\mbox{I}^{\rho}_{\mbox{\scriptsize n}}-1/2(1-r_0')]/r_0'}
    {r_0'[1/2(1+r_0')-\mbox{I}^{\rho}_{\mbox{\scriptsize max}}/\mbox{I}^{\rho}_{\mbox{\scriptsize n}}]+
    [\mbox{I}^{\rho}_{\mbox{\scriptsize min}}/\mbox{I}^{\rho}_{\mbox{\scriptsize n}}-1/2(1-r_0')]/r_0'}}
\end{eqnarray}
\end{widetext}
with a normalization factor $\mbox{I}^{\rho}_{\mbox{\scriptsize n}}=2\mbox{I}^{\rho}_0/(1+r_0')$.
I$^{\rho}_0$ is the intensity measured at U$_\phi=\1$.
\newline\indent The noncyclic geometric phase is given by $\phi_g=-\Omega/2$,
where $\Omega$ is the solid angle enclosed by a geodesic path and its
shortest geodesic closure on the Bloch sphere 
\cite{SamuelBhandari1988}: $\phi_{\mbox{\scriptsize{g}}}$ 
and the total phase $\phi$ 
are related to the path by the polar and azimuthal 
angles $2\xi$ and $2\delta$,
so that the pure state geometric phase becomes
\begin{eqnarray}\label{eq:gamma(delta,xi)}
    \phi_{\mbox{\scriptsize{g}}}
=\phi-
\phi_{\mbox{\scriptsize{d}}}=\delta[1-\cos{(2\xi)}],
\end{eqnarray}
while its dynamical counterpart is
\begin{eqnarray}\label{eq:delta(phi,xi)}
    \phi_{\mbox{\scriptsize{d}}}=\delta\cos{(2\xi)}.
\end{eqnarray}
By proper choice of $2\xi$ and $2\delta$, 
U$_\phi$ can be set to generate
purely geometric, purely dynamical, or arbitrary
combinations of both phases as shown in Fig.\,\ref{fig2}.

The theoretical prediction for the mixed state phase is \cite{SjoeqvistEtAl2000,LarssonSjoeqvist2003}
\begin{eqnarray}\label{eq:MixedStatePhase}
    \Phi(r_0')&=&\arctan\left(r_0'\tan\delta\right)
\end{eqnarray}
To access Eq.\,(\ref{eq:MixedStatePhase})
experimentally $r_0'$ needs to be varied.
In addition to the DC current, which effects the 
transformation U$_1$, random noise is applied to the first coil, thereby
changing B$_x$ in time. Neutrons, which are part of the 
ensemble
$\rho_{\mbox{\scriptsize{in}}}(r_0')$, arrive at different 
times at the coil and experience different magnetic field 
strengths. This is
equivalent to applying different unitary operators
U$(\pi/4\!+\!\Delta\xi'(t),0,-\pi/2)=\mbox{\~U}_1(\Delta\xi'(t))$. For the
whole ensemble we have to take the time integral 
\begin{eqnarray}
    \rho=\int\mbox{\~U}_1(\Delta\xi'(t))|+\rangle\langle +|\mbox{\~U}_1^{\dagger}(\Delta\xi'(t))dt\nonumber.
\end{eqnarray}
Although each separate transformation is
unitary, due to the randomness of the signal we end up 
with a
nonunitary evolution that yields a mixed state
\cite{BertlmannEtAl2006}. 
Note that in this method the purity $r_0'$ of the input 
state 
is not affected before \~U$_1$ creates spin superpositions 
distributed around $|\psi_0\rangle$ within the $y,z$-plane. 
We are left with $\vec r_0\!=\!\left(0,-r_0,0\right)$
where $r_0\!<\!1$, as has been confirmed by a 3D spin
analysis of the state $\mbox{\~U}_1\rho_{\mbox{\scriptsize{in}}}
\mbox{\~U}^{\dagger}_1$.
\begin{figure}
    \scalebox{0.485}{\includegraphics{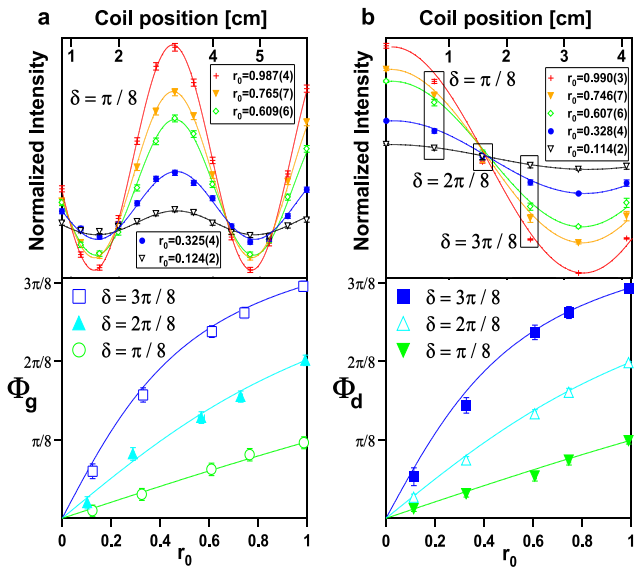}}
    \caption{Above: typical measured intensities for five 
values of purity $r_0$ versus coil positions. 
	Below: resulting mixed state phases 
$\Phi_{\mbox{\scriptsize{g}}}$ and 
$\Phi_{\mbox{\scriptsize{d}}}$ versus purity $r_0$.
	a) Purely geometric case: $2\xi\!=\!\pi/2$. 
Data indicated by circles is calculated from oscillations 
above.
    b) Purely dynamical case: $2\xi\!=\!0$. Only one 
intensity value - marked by rectangles in the above plot - 
is needed for each $\Phi_{\mbox{\scriptsize{d}}}$ (see text). }
    \label{fig3}
\end{figure}

\begin{figure}
    \scalebox{0.4}{\includegraphics{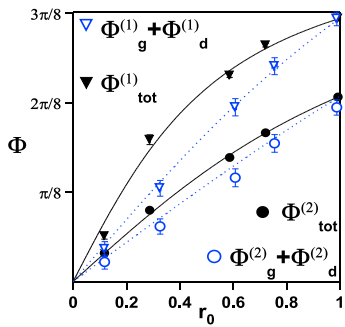}}
    \caption{Filled markers: Measured total 
mixed state phase $\Phi_{\mbox{\scriptsize{tot}}}$ 
versus purity $r_0$ for two examples of 
U$_{\mbox{\scriptsize{tot}}}$ associated to the total pure state phases $\phi_{\mbox{\scriptsize{g}}}^{(1,2)}
+\phi_{\mbox{\scriptsize{d}}}^{(1,2)}$ (see text).
Open markers: 
$\Phi_{\mbox{\scriptsize{g}}}^{(1,2)}
+\Phi_{\mbox{\scriptsize{d}}}^{(1,2)}$ 
as calculated 
from measured data in Fig. \ref{fig3}.	
The solid and dotted lines are theory
curves assuming either nonadditivity or additivity, 
respectively. 
The data clearly demonstrate the nonadditivity of the 
Sj\"{o}qvist mixed state phase.}
    \label{fig4}
\end{figure}

The experiment was carried out at the 
research reactor facility of the Vienna University of 
Technology. 
A neutron beam -- incident from a pyrolytic graphite 
crystal -- with a mean wavelength 
$\lambda\!\approx\!1.98$\,\AA ~and spectral width $\Delta\lambda/\lambda\!\approx\!0.015$, was polarized 
up to $r_0'\!\approx\!99$\,\% by reflection from a bent 
Co-Ti supermirror array. 
The analyzing supermirror was slightly de-adjusted to
higher incident angles to suppress second order harmonics in the
incident beam. The final maximum intensity was about 150\,counts/s
at a beam cross-section of roughly 1\,cm$^2$. A $^3$He gas detector with high efficiency for neutrons of this energy range 
was used. Spin rotations around the $+x$-axis were implemented by
magnetic fields B$_x$ from DC coils made of anodized aluminium
wire (0.6\,mm in diameter) wound on frames with rectangular
profile ($7\!\times\!7\!\times\!2$ cm$^3$). Coil windings in
$z$-direction provided for compensation of the guide field 
B$_z$ at coil positions (see Fig.\,\ref{fig1}) for B$_x$ 
to be parallel to the $x$-direction. 
B$_z$ was realized
by two rectangular coils of 150\,cm length in Helmholtz geometry.
Maximum coil currents of 2\,A and guide field strengths of 1\,mT
were sufficient to achieve required spin rotations and prevent unwanted
depolarization. The noise from a standard signal
generator consisted of random DC offsets varying at a rate
of 20\,kHz. 
In order to find the coil current values for 
required spin rotation angles, each coil current was scanned
separately. A measured intensity minimum after stepwise 
increase indicates that a $\pi$ rotation has been achieved. 
Then, with both coils mounted on translation stages and set 
to a $\pi/2$ rotation around the $x$-axis, inter-coil 
distances were varied to 
search for intensity minima.
The distance between two minima (6-7
cm in our case) corresponds to a $2\pi$ rotation around B$_z$
within the $x,y$
plane.
\newline\indent For purely geometric phases the parameter sets
($\xi\!=\!\pi/4,\delta,\zeta\!=\!\delta\!-\!\pi/2$) with
$\delta\!=\!\phi_{\mbox{\scriptsize{g}}}\!
=\!\pi/8,2\pi/8,3\pi/8$ were chosen.
For each set the intensity oscillations I$^\rho$ (see Fig.\,\ref{fig3}a, upper graph; 
data shown for $\delta=\phi_{\mbox{\scriptsize{g}}}=\pi/8$) were
measured, scanning the position of the second coil for five
values of $r_0$. 
For the purely dynamical phase - since
$2\xi=0$ and Eq.\,(\ref{mixedstateintensity}) reduces to
I$^\rho=(1\!-\!r_0')/2\!+\!r_0'\cos^2\delta$ - only one intensity value is needed. 
With the second coil turned
off, the distance between the
first and the third coil was chosen such that
$\delta=\phi_{\mbox{\scriptsize{d}}}=\pi/8,2\pi/8,3\pi/8$ for five values of $r_0$ (Fig.\,\ref{fig3}b, upper graph). 
By Eq.\,(\ref{eq:MixedStatePhInTermsOfImaxImin}) the 
geometric and
dynamical mixed state phases 
$\Phi_{\mbox{\scriptsize{g}}}(r_0)$ and 
$\Phi_{\mbox{\scriptsize{d}}}(r_0)$ (Fig.\,\ref{fig3}, lower graphs) were
calculated from intensity values extracted from 
least square fits for I$^\rho$ (solid lines in 
Fig.\,\ref{fig3}, upper graphs). 
Vertical error bars in the lower graphs contain two about 
equal contributions: fitting errors and an 
estimated uncertainty of 0.5 mm for the reproduction of coil 
positions. 
The small horizontal error bars stem from the statistical 
uncertainty 
of the purity measurements. 
The solid lines are theoretical curves according to Eq.\,(\ref{eq:MixedStatePhase}), using the measured value 
for $\Phi$ without noise as phase reference. 
The experimental data reproduce well the $r_0$-dependence 
predicted by Eq.\,(\ref{eq:MixedStatePhase}). 
\linebreak\indent 
We want to emphasize that our investigation focuses on a 
special property of the mixed state phase: 
its nonadditivity. Since the Sj\"{o}qvist phase is defined as a weighted average of phase
factors rather then one of phases (see
\cite{FuChen2004a,SinghEtAl2003} for a more elaborate discussion)
it is true only for pure states
that phases of separate measurements can be
added up to a total phase. 
Suppose we carry
out two measurements on a pure state system:
the state is subjected to a unitary transformation 
U$_{\mbox{\scriptsize{g}}}$ in the first and to
U$_{\mbox{\scriptsize{d}}}$ in the second measurement 
inducing the phases $\phi_{\mbox{\scriptsize{g}}}$ and $\phi_{\mbox{\scriptsize{d}}}$, respectively. 
Applying Eqs.\,(\ref{eq:gamma(delta,xi)}) and
(\ref{eq:delta(phi,xi)}), we can also choose a combination 
of angles
$2\xi$ and $2\delta$ leading to a transformation 
U$_{\mbox{\scriptsize{tot}}}$ 
so that we measure the total pure state phase 
$\phi_{\mbox{\scriptsize{g}}}+\phi_{\mbox{\scriptsize{d}}}$
(note that the three evolution paths induced by U$_{\mbox{\scriptsize{g}}}$, U$_{\mbox{\scriptsize{d}}}$ and U$_{\mbox{\scriptsize{tot}}}$ differ from each other).
However, the result of the latter experiment
for the system in a mixed state is
$\Phi_{\mbox{\scriptsize{tot}}}(r_0)
=\arctan{\left[r_0\tan(\phi_{\mbox{\scriptsize{g}}}
+\phi_{\mbox{\scriptsize{d}}})\right]}$. 
The total phase is then \emph{not} given by 
$\Phi_{\mbox{\scriptsize{g}}}(r_0)
+\Phi_{\mbox{\scriptsize{d}}}(r_0)$, with
$\Phi_{\mbox{\scriptsize{g}}}(r_0)
=\arctan{(r_0\tan\phi_{\mbox{\scriptsize{g}}})}$ and $\Phi_{\mbox{\scriptsize{d}}}(r_0)
=\arctan{(r_0\tan\phi_{\mbox{\scriptsize{d}}})}$.
In our experiment we have chosen two examples of 
U$_{\mbox{\scriptsize{tot}}}$, i.e. two sets of values for 
B$_x$ in
the second coil 
($2\xi^{(1)}\!=\!60^{\circ},~2\xi^{(2)}\!=\!48^{\circ}$) and
the distance within B$_z$ 
($2\delta^{(1)}\!=\!90^{\circ},~2\delta^{(2)}\!
=\!135^{\circ}$). 
According to Eqs.\,(\ref{eq:gamma(delta,xi)}) and
(\ref{eq:delta(phi,xi)}) the total pure state phases 
$\phi_{\mbox{\scriptsize{g}}}^{(1)}
+\phi_{\mbox{\scriptsize{d}}}^{(1)}$ and 
$\phi_{\mbox{\scriptsize{g}}}^{(2)}
+\phi_{\mbox{\scriptsize{d}}}^{(2)}$ with
$\phi^{(1)}_{\mbox{\scriptsize{g}}}
=\phi^{(2)}_{\mbox{\scriptsize{g}}}=\pi/8$ and 
$\phi^{(1,2)}_{\mbox{\scriptsize{d}}}=2\pi/8,\pi/8$ 
are obtained from intensity oscillations. 
In Fig.\,\ref{fig4} the resulting mixed state 
phases $\Phi^{(1,2)}_{\mbox{\scriptsize{tot}}}$ 
and the sum $\Phi_{\mbox{\scriptsize{g}}}^{(1,2)}
+\Phi_{\mbox{\scriptsize{d}}}^{(1,2)}$ are plotted. 
\indent Note that this nonadditivity of mixed state phases is 
not due to the nonlinearity 
of the geometric phase, that occurs -- for instance -- when 
the system evolves close to the orthogonal state of the 
reference state \cite{Bhandari1997}. 
\newline\indent Recently there has been a report on NMR experiments
\cite{DuEtAl2007} investigating Uhlmann's mixed state geometric
phase. It is a property of a composite system  
undergoing a certain non-local evolution of system and ancilla \cite{EricssonEtAl2003}. 
Diverse phase definitions, depending on this evolution, are 
possible. The phase investigated in the present paper is a 
special case in which
the ancilla does not necessarily evolve. 
While the preconditions for inherent fault 
tolerance remain intact for the 
Sj\"{o}qvist phase, the question whether 
other phases offer advantages in
terms of robustness remains an exciting issue of discussion. 
\linebreak\indent To summarize, we have measured spin-1/2 
mixed state phases with polarized neutrons.
Their dependence on the purity of the input state was 
observed. 
Our results show that
these phases are not additive. The sum of phases
measured in separate experiments is not equal to the result 
of one single measurement, in contrast to what could 
be expected from straightforward extrapolation 
of pure state behavior. This interesting
property of the Sj\"{o}qvist mixed state phase might be of high relevance for possible applications of geometric phases. 
\linebreak\indent We thank E. Balcar for critical reading 
of the manuscript. 
This work was
financed by the Japan Science and Technology Agency (JST) and the
Austrian Science Fund (FWF, Project. Nr. P 17803-N02).

\end{document}